\documentclass[prd,nofootinbib,preprint,superscriptaddress]{revtex4}

\usepackage{amsmath, amssymb, amsthm, graphicx, epsfig, fancyhdr,epsfig, slashed, mathrsfs}
\usepackage[T1]{fontenc}
\usepackage{tikzsymbols}
\usepackage{natbib}
\usepackage{float}
\usepackage[thinc]{esdiff}
\usepackage{feynmf}
\usepackage{tikzsymbols}
\usepackage{natbib}
\usepackage{ amssymb }
\usepackage{amsmath}
\usepackage{dsfont}
\usepackage{cancel}
\usepackage[titletoc]{appendix}
\usepackage{float}
\usepackage{bm}

\usepackage{tikz,xcolor,hyperref}

\usepackage{slashed}

\definecolor{lime}{HTML}{A6CE39}
\DeclareRobustCommand{\orcidicon}{
	\begin{tikzpicture}
	\draw[lime, fill=lime] (0,0) 
	circle [radius=0.2] 
	node[white] {{\fontfamily{qag}\selectfont \tiny ID}};
	\draw[white, fill=white] (-0.0625,0.095) 
	circle [radius=0.007];
	\end{tikzpicture}
	\hspace{-2mm}
}

\foreach \x in {A, ..., Z}{\expandafter\xdef\csname orcid\x\endcsname{\noexpand\href{https://orcid.org/\csname orcidauthor\x\endcsname}
			{\noexpand\orcidicon}}
}

\usepackage{tikz,xcolor,hyperref}

\definecolor{lime}{HTML}{A6CE39}
\DeclareRobustCommand{\orcidicon}{
	\begin{tikzpicture}
	\draw[lime, fill=lime] (0,0) 
	circle [radius=0.2] 
	node[white] {{\fontfamily{qag}\selectfont \tiny ID}};
	\draw[white, fill=white] (-0.0625,0.095) 
	circle [radius=0.007];
	\end{tikzpicture}
	\hspace{-2mm}
}

\foreach \x in {A, ..., Z}{\expandafter\xdef\csname orcid\x\endcsname{\noexpand\href{https://orcid.org/\csname orcidauthor\x\endcsname}
			{\noexpand\orcidicon}}
}

\newcommand{\be}{\begin{equation}}
\newcommand{\ee}{\end{equation}}
\newcommand{\bea}{\begin{eqnarray}}
\newcommand{\eea}{\end{eqnarray}}

\usepackage{amsmath}
\usepackage{braket}
\usepackage{graphicx}
\usepackage{mathrsfs}
\usepackage{xcolor}
\usepackage{float}
\usepackage{gensymb}
\usepackage{amssymb}
\usepackage{slashed}
\usepackage{latexsym}
\usepackage{breqn}
\usepackage{subfig}

\newcommand{\ba}{\begin{eqnarray}}
\newcommand{\ea}{\end{eqnarray}}
\newcommand{\bi}{\begin{itemize}}
\newcommand{\ei}{\end{itemize}}

\newcommand{\x}{\star}

\newcommand{\arXivold}[2]{\href{http://arxiv.org/pdf/#1}{{\tt #2/#1}}}
\newcommand{\arXiv}[2]{\href{http://arxiv.org/pdf/hep-ph/#1}{{\tt #2/#1}}}

\begin{document}

\title{Phenomenological Aspects of Lee-Wick QED}

\author{Fayez Abu-Ajamieh\orcidA{}}
\email{fayezajamieh@iisc.ac.in}
\affiliation{Centre for High Energy Physics; Indian Institute of Science; Bangalore; India}

\author{Pratik Chattopadhyay\orcidB{}}
\email{pratikpc@gmail.com}
\affiliation{Tsung-Dao Lee Institute; Shanghai Jiao Tong University; Shanghai; China}
\affiliation{School of Physics; The University of Electronic Science and Technology of China; 88 Tian-run Road; Chengdu; China}

\author{Marco Frasca\orcidC{}}
\email{mar.frasca@gmail.com}
\affiliation{Rome; Italy}

\begin{abstract}
We study some phenomenological aspects of Lee-Wick (LW) QED. In particular, we show that LW QED implies charge dequantization and a flavor-dependent LW scale. We study the implications of the Weak Gravity Conjecture (WGC) in LW QED and calculate the modified electric force and potential and use the former to reformulate the WGC in LW QED. We also calculate the photon self-energy and the Uehling potential in LW QED. We show that bounds on milli-charged particles from matter neutrality experiments set stringent limits on the LW scale of fermions and of the photon.
\end{abstract}
\maketitle

\section{Introduction}\label{sec:1}
Studying Quantum Field Theories (QFTs) with non-renormalizable high-order derivatives has a long history. One particular example was studied by Lee and Wick \cite{Lee:1969fy, Lee:1970iw}, where they assumed that the Pauli-Villars regulator in QED corresponds to a physical particle.  Such a theory has non-renormalizable higher-order derivative operators and can be shown to have two poles, the massless photon and the massive LW photon. The massive LW photon has a problem in that its residue has the wrong sign, which leads to the theory being acausal \cite{Coleman}. Nonetheless, it can be shown that the unitarity of the $S$ matrix can be preserved via deformation of the integration contour and as long as the massive LW photon decays quickly \cite{Lee:1969fy, Lee:1970iw, Cutkosky:1969fq}, thereby making sense of the theory.

As an example to clarify LW theories, consider the following toy model presented in \cite{Grinstein:2007mp} 
\begin{equation}\label{eq:ToyModel1}
    \mathcal{L} = \frac{1}{2}\partial_{\mu}\hat{\phi} \partial^{\mu}\hat{\phi} -\frac{1}{2}m^{2}\hat{\phi}^{2} - \frac{1}{M^{2}} (\partial^{2}\hat{\phi})^{2},
\end{equation}
where $M$ is the LW scale. It is easy to show that the propagator is given by
\begin{equation}\label{eq:LW_propagator}
    D_{\text{LW}} = \frac{i}{p^{2}-m^{2}-p^{4}/M^{2}},
\end{equation}
and we can see that the propagator has two poles. For $M \gg m$, these poles are given by $p^{2} \simeq m^{2}, M^{2}$. Using the auxiliary field formalism described in \cite{Grinstein:2007mp} to isolate the LW degree of freedom, it can be shown that eq. (\ref{eq:ToyModel1}) can be reformulated as
\begin{equation}\label{eq:ToyModel2}
    \mathcal{L} = \frac{1}{2}\partial_{\mu}\phi\partial^{\mu}\phi -  \frac{1}{2}\partial_{\mu}\tilde{\phi}\partial^{\mu}\tilde{\phi} + \frac{1}{2}M^{2} \tilde{\phi}^{2} -\frac{1}{2}m^{2}(\phi - \tilde{\phi})^{2},
\end{equation}
where $\phi = \hat{\phi} + \tilde{\phi}$ and $\tilde{\phi}$ is the auxiliary field, and in this formalism, we can see the two fields explicitly. Notice that the propagator of the auxiliary field is given by
\begin{equation}
    \tilde{D} = \frac{-i}{p^{2} - M^{2}},
\end{equation}
and we see that this propagator has the wrong sign, implying that $\tilde{\phi}$ corresponds to a ghost degree of freedom. It is this sign that is the source of acausality in LW models.

Lee-Wick theories have been suggested as a solution to the hierarchy problem in \cite{Grinstein:2007mp}. They also have been considered in a non-perturbative regime \cite{Frasca:2022duz} where confinement was shown in the non-Abelian case, similar to what is observed in the ordinary case. Furthermore, it was shown that the theory develops a mass gap. There also has been considerable progress in adopting Lee Wick models in quantum gravity \cite{Modesto:2016ofr}. In fact, it has been shown that such theories can be either super-renormalizable or UV finite.

This paper is organized as follows: In Section \ref{sec:2} we briefly review LW QED. In Section \ref{sec:3} we discuss electric charge dequantization in LW QED and discuss the possibility of LW QED solving the $(g-2)_{\mu}$ anomaly. In Section \ref{sec:4} we discuss the implications of the WGC in LW QED. In Section \ref{sec:5} we derive the electric force and potential in LW QED and compare them to their Coulomb counterparts. In Section \ref{sec:6} we calculate the photon self-energy in LW QED and use it to calculate the LW Uehling potential and Lamb shift and finally we conclude in Section \ref{sec:7}

\section{Review of Lee-Wick QED}\label{sec:2}
To discuss LW QED, it is convenient to use the formalism introduced in \cite{Grinstein:2007mp}. In this formalism, the authors have written down the fermion Lagrangian for a single left-handed quark doublet. In our case, we will be dealing with Dirac fermions. Thus, a straightforward generalization of the left-handed doublet to a Dirac fermion yields the following LW Lagrangian for QED
\begin{equation}
\label{LW}
    \mathcal{L}_{\text{LW}}=-\frac{1}{4}F_{\mu\nu}F^{\mu\nu}+\frac{1}{M^2_A}D^\mu F_{\mu\nu}D^\lambda F_{\lambda}^{\nu}+{\bar\Psi}\left(i{\slashed D}+\frac{i}{M_Q^2}{\slashed D}{\slashed D}{\slashed D}-m\right)\Psi,
\end{equation}
where $M_A$ and $M_Q$ are the LW scales of the photon and the fermion respectively, and the covariant derivative is given by $D_{\mu} = \partial_{\mu} + i e Q A_{\mu}$, and the Lagrangian in Eq. (\ref{LW}) is invariant under $U(1)$. It is clear from the above expression that the second term in the Lagrangian is the higher derivative term for the photon and the fourth term is the higher derivative term for the Dirac fermion. Our generalization from a left-handed quark doublet to a Dirac fermion doesn't necessarily lead to any inconsistencies. The only modification it does is to add a mass term for the fermion.
The Feynman rules corresponding to the propagators and interaction vertex for the above Lagrangian are given by
\begin{eqnarray}
\Pi_{g}^{\mu\nu}(p) & = & \frac{iM^2_A}{p^{2}(p^2-M^2_A)}
\left(g^{\mu\nu}-(1-\xi)\frac{p^\mu p^\nu}{p^2}-\xi\frac{p^\mu p^\nu}{M^2_A}\right), \label{eq:photon_propagator} \\
\Pi_{f}(p) & = & \frac{i[(1+p^2/M_Q^2)\slashed{p}+m]}{(1+p^2/M^2_Q)^2p^{2}-m^{2}},
\label{eq:fermion_propagator}\\
V^{\mu}(q_{1},q_{2}) & = & -ieQ\gamma^\mu+i\frac{eQ}{M^2_Q}\left(q_1^2\gamma^\mu+{\slashed q}_2q_1^\mu
-\frac{1}{2}({\slashed p}\gamma^\mu-\gamma^\mu{\slashed p}){\slashed q}_2
\right), \label{eq:vertex}
\end{eqnarray}
where $q_{1,2}$ are the momenta of the (outgoing) fermions, and we can see that in the limit $M_A, M_A \rightarrow \infty$, the Feynman rules of ordinary QED are retrieved. The above formulation of the LW QED is quite suitable for computing amplitudes at tree level, however, calculating loop diagrams using this formalism can be quite complex. Therefore, loop calculation is much easier done using the auxiliary field formalism described below. Using this formalism, the fermionic part of the Lagrangian can be expressed as
\begin{align}
\label{nlw}
\mathcal{L}_{AF}={\bar\Psi}(i{\slashed D-m})\Psi
    +i\bar{\tilde{\Psi}}{\slashed D}\tilde\Psi
    +\bar{\Psi}i\slashed{D}\tilde\Psi+\bar{\hat{\Psi}} i\slashed{D}\hat{\Psi}
+M_Q\bar{\tilde\Psi}\tilde{\Psi},
\end{align}
where $\tilde{\Psi}$ and $\hat{\Psi}$ are two auxiliary fermion doublets. Using their equation of motion, 
\begin{align}
\hat{\Psi}& = -\frac{i\slashed{D}}{M_Q}\Psi,\\
\tilde{\Psi} & =-\frac{\slashed{D}\slashed{D}}{M_Q^2}\Psi,
\end{align}
and putting them back to the Lagrangian in eq. (\ref{nlw}), we obtain the exact fermion terms in the higher derivative Lagrangian in eq. (\ref{LW}). We now use the shift $\Psi=\Psi'-\tilde{\Psi}$ to rewrite the above Lagrangian in eq. (\ref{nlw}) in an explicit form
\begin{align}
   \mathcal{L}_{\text{AF}}={\bar\Psi'}(i{\slashed D-m})\Psi'
    +\bar{\tilde{\Psi}}\left(i{\slashed D+m}\right)\Psi
    +m(\bar{\Psi'}\tilde{\Psi}+\bar{\tilde{\Psi}}\Psi')
    +\bar{\hat{\Psi}} i\slashed{D}\hat{\Psi}
+M_Q\bar{\tilde\Psi}\tilde{\Psi}.
\end{align} 

In our calculation, we take the LW scale $M_Q$ to be much greater than the mass of the fermion $m$ and ignore the mass terms for LW fermions. This yields
\begin{align}
   \mathcal{L}_{\text{AF}}={\bar\Psi'}(i{\slashed D-m})\Psi'
    +\bar{\tilde{\Psi}}i\slashed {D}\tilde{\Psi}
    +\bar{\hat{\Psi}} i\slashed{D}\hat{\Psi}
+M_Q\bar{\tilde\Psi}\hat{\Psi},
\end{align}  
and we can see that after separating the kinetic terms explicitly, the mass term is no longer diagonal. To diagonalize it, we use transformation of the fields 
\begin{equation}
\begin{pmatrix}
\tilde{\Psi}\\
\hat{\Psi} 
\end{pmatrix}
=\begin{pmatrix}
\cos\theta & \sin\theta\\
-\sin\theta & \cos\theta
\end{pmatrix}
\begin{pmatrix}
\tilde{\Psi'}\\
\hat{\Psi'} 
\end{pmatrix}.
\end{equation}

This transformation leaves the derivative terms unchanged, but diagonalizes the mass term if we choose $\cos^2\theta - \sin^2\theta=0$. Thus we set $\sin\theta =\frac{1}{\sqrt{2}}$, and $\cos\theta =\frac{1}{\sqrt{2}}$. For simplicity, we write the complete diagonalized Lagrangian without prime notation, i.e we set $\tilde{\Psi'}=\tilde{\Psi}$ and $\hat{\Psi'}=\hat{\Psi}$. The Lagrangian reads
\begin{align}\label{eq:QED_Lag_LW_final}
   \mathcal{L}_{AF}={\bar\Psi'}(i{\slashed D-m})\Psi'
    +\bar{\tilde{\Psi}}(i\slashed {D}-M_Q)\tilde{\Psi}
    +\bar{\hat{\Psi}} (i\slashed{D}+M_Q)\hat{\Psi}.
\end{align}  
This is the Lagrangian that will be used to perform loop calculations. As we can see, this Lagrangian lacks any higher derivatives, however, it has two extra fermionic (LW) fields. This will make our computation much easier. The Feynman rules can be easily extracted from eq. (\ref{eq:QED_Lag_LW_final}). We provide them explicitly in the Appendix.

\section{Charge Dequantization and the $g-2$ Anomaly}\label{sec:3}
Although in the minimal SM the electric charge is quantized, charge dequantization can be introduced to the SM in several ways, such as by adding an $SU(2) \times SU(2)$ Dirac fermion with hypercharge $2\epsilon$ \cite{Okun:1983vw}, introducing an additional $U(1)$ gauge group that mixes with the SM $U(1)_{\text{QED}}$ \cite{Holdom:1985ag}, allowing neutrinos to have a charge \cite{Foot:1990uf, Foot:1992ui}, or through the introduction of non-locality \cite{Capolupo:2022awe, Abu-Ajamieh:2023roj, Chattopadhyay:2023nbj}. Here, we show that the electric charge is also dequantized in LW QED.

But before we do so, we should point out that a similar calculation was attempted in \cite{Turcati:2014eaa, Turcati:2016aog}, however, in their calculation, they assumed that charge is quantized in LW QED (i.e., they have set $F_{1} = 1$ without calculation). In addition, in their calculation of the $g-2$, they assumed that the interaction vertex is the same as the SM, and that the fermionic sector remains unaltered. Furthermore, they assumed that the tree level calculation in LW QED remains the same as in the SM case, which as we shall show below, is not case. We believe that their treatment is inadequate and below we show the proper calculation below.

To this avail, we calculate the form factors $F_{1}(p^{2})$ and $F_{2}(p^{2})$ from the $\gamma \overline{f}f$ vertex. Usually one has to calculate the 1-loop correction to the vertex. Luckily, here we don't have to calculate any loop corrections, as in LW QED, $F_{1}(p^{2})$ and $F_{2}(p^{2})$ are already modified compared to their SM values at tree level. Given the LW Feynman rules (eqs. \ref{eq:photon_propagator} - \ref{eq:vertex}), and using the standard procedure of Dirac algebra and the Gordon identity, we can express the $\gamma \overline{f}f$ vertex as 
\begin{equation}\label{eq:Aff_vertex1}
    i \mathcal{M}^{\mu} = (-i e) \overline{u}(q_{2})\Big[\Big(1-\frac{2m^{2}}{M_Q^{2}} \Big)\gamma^{\mu} + \frac{i \sigma^{\mu\nu}}{2m} p_{\nu} \Big( \frac{3m^{2}}{M^{2}_Q}\Big)\Big] u(q_{1}).
\end{equation}
Comparing this to the canonical form
\begin{equation}\label{eq:Canonical_form)}
     i \mathcal{M}^{\mu} = (-i e) \overline{u}(q_{2})\Big[F_{1}(p^{2}) \gamma^{\mu} + \frac{i \sigma^{\mu\nu}}{2m} p_{\nu} F_{2}(p^{2})\Big] u(q_{1}),
\end{equation}
we immediately identify\footnote{Notice that strictly speaking, $m$ appearing in eqs (\ref{eq:f1}) and  (\ref{eq:f2}) is actually the bare mass $m_{0}$, however, $m \simeq m_{0}$ at the leading order in $(m^{2}/M_{Q}^{2})$. To see this, the physical mass can be obtained from the fermion propagator in eq. (\ref{eq:fermion_propagator}) by requiring the on-shell condition $\Big( 1+ \frac{\slashed{p}\slashed{p}}{M_{Q}^{2}}\Big)\slashed{p}-m_{0}\Big|_{\slashed{p}=m} = 0$, which implies that $m_{0} = m \Big(1+\frac{m^{2}}{M_{Q}^{2}}\Big)$. Thus, for $m \ll M_{Q}$, $m_{0} \simeq m$ at LO.}
\begin{align}
    F_{1}(p^{2}) & = 1 - \frac{2m^{2}}{M^{2}_Q}, \label{eq:f1}\\\
    F_{2}(p^{2}) & = \frac{3m^{2}}{M^{2}_Q}. \label{eq:f2}
\end{align}

Notice that since $F_{1}(0)$ essentially renormalizes the electric charge, we see from eq. (\ref{eq:f1}) that introducing the LW higher-order operator leads the electric charge to deviate from unity. We can also see that when we take $M_Q \rightarrow \infty$, the SM charge quantization condition $ F_{1}(0) = 1$ is restored as expected. Eq. (\ref{eq:f1}) shows that the deviation is dependent on the \textit{ratio} $m/M_Q$. This implies that there are two options for charge dequantization: (1) Either the LW scale $M_Q$ is universal, implying a flavor-dependant electric charge, or (2) the electric charge is universal, implying a flavor-dependent LW scale.

Let's evaluate these two options. It can be shown that if neutrinos are Dirac fermions, then the requirement of anomaly cancellation means that the only way charge can be dequantized is as follows (see for instance \cite{Foot:1990uf,Foot:1992ui})
\begin{align}
    Q_{\nu_{L}}^{i} = Q_{\nu_{R}}^{i} & = \epsilon, \label{eq:q_deq_1} \\
    Q_{e_{L}}^{i} = Q_{e_{R}}^{i} & = -1 + \epsilon,  \label{eq:q_deq_2} \\
    Q_{u_{L}}^{i} = Q_{u_{R}}^{i} & = \frac{2}{3} + \frac{1}{3}\epsilon,  \label{eq:q_deq_3}\\
    Q_{d_{L}}^{i} = Q_{d_{R}}^{i} & = -\frac{1}{3} +\frac{1}{3}\epsilon,  \label{eq:q_deq_4}
\end{align}
where $i = 1,2,3$ is the generation index. It is quite easy to see that eqs. (\ref{eq:q_deq_1}) - (\ref{eq:q_deq_4}) can never be satisfied if the electric charge is flavor-dependent. Thus, we are only left with the option of a flavor-dependent LW scale. Comparing eqs. (\ref{eq:q_deq_1}) - (\ref{eq:q_deq_4}) with eq. (\ref{eq:f1}), it is quite easy to see that
\begin{equation}\label{eq:epsilon}
    \epsilon = \frac{2m_{l}^{2}}{M_{l}^{2}} = \frac{4m_{u}^{2}}{M_{u}^{2}} = \frac{2m_{d}^{2}}{M_{d}^{2}}.
\end{equation}

On the other hand, if neutrinos are Majorana fermions, then the neutrality of the Weinberg operator (in addition to the anomaly cancellation requirement), implies that they must be neutral \cite{Babu:1989tq, Babu:1989ex, Foot:1992ui}. This means that LW QED is inconsistent with Majorana neutrinos, which furnishes a suitable test for the theory from experiments related to the nature of neutrinos

Assuming that neutrinos are Dirac fermions, we can use eq. (\ref{eq:epsilon}) to set bounds on the (flavor-dependent) LW scale by utilizing the experimental bounds on the charge of neutrinos. The most stringent bounds on the charge of neutrinos arise from matter neutrality \cite{Foot:1992ui, Giunti:2014ixa, Raffelt:1999gv}
\begin{equation}\label{eq:Nu_Q_limit}
    |\epsilon| < 10^{-21},
\end{equation}
which translates to the following bounds
\begin{eqnarray}
M_{e} & \gtrsim & 2.29 \times 10^{4} \hspace{2mm} \text{TeV}, \label{eq:e_bound}\\
M_{\mu} & \gtrsim & 4.74 \times 10^{6} \hspace{2mm} \text{TeV}, \label{eq:mu_bound}\\
M_{\tau} & \gtrsim & 7.95 \times 10^{7} \hspace{2mm} \text{TeV}, \label{eq:tau_bound}\\
M_{u} & \gtrsim & 1.37 \times 10^{5} \hspace{2mm} \text{TeV}, \label{eq:u_bound}\\
M_{d} & \gtrsim & 2.09 \times 10^{5} \hspace{2mm} \text{TeV}, \label{eq:d_bound}\\
M_{s} & \gtrsim & 4.16 \times 10^{6} \hspace{2mm} \text{TeV}, \label{eq:s_bound}\\
M_{c} & \gtrsim & 8.03 \times 10^{7} \hspace{2mm} \text{TeV}, \label{eq:c_bound}\\
M_{b} & \gtrsim & 1.87 \times 10^{8} \hspace{2mm} \text{TeV}, \label{eq:b_bound}\\
M_{t} & \gtrsim & 1.09 \times 10^{10} \hspace{2mm} \text{TeV}. \label{eq:t_bound}
\end{eqnarray}

Next, we turn our attention to the modification of the second form factor $F_{2}(p^{2})$. The SM value of $F_{2}(0) = 0$ at tree-level, which we see is also retrieved in the limit $M \rightarrow \infty$. Since $F_{2}(0)$ is the contribution to the magnetic dipole moment, we see that LW QED could in theory provide a solution to the $(g-2)_{\mu}$ anomaly
\begin{eqnarray}\label{eq:LW_g-2}
    (g-2)_{\mu} = 2F_{2}(0) = \frac{6m_{\mu}^{2}}{M_{\mu}^{2}}.
\end{eqnarray}

It is known that there is a discrepancy between the theoretically calculated (see \cite{Aoyama:2020ynm} and the references therein) and experimentally measured \cite{Muong-2:2006rrc, Muong-2:2021ojo, Muong-2:2021ovs, Muong-2:2021vma, Muong-2:2023cdq} magnetic dipole moment of the muon. This discrepancy currently stands at $5.0\sigma$:\footnote{We should point out that recent high-precision lattice QCD simulations \cite{Borsanyi:2020mff, Ce:2022kxy,ExtendedTwistedMass:2022jpw} appear to agree with the experimental results. As discussed in \cite{Abu-Ajamieh:2022nmt, Abu-Ajamieh:2023qvh}, these results are in tension with the data-driven approach, and that has stirred a debate recently as to whether the anomaly is genuine or not. We ignore this tension in this paper.}
\begin{equation}\label{eq:g-2_measurement}
\Delta a_{\mu} = a_{\mu}^{\text{Exp}} - a_{\mu}^{\text{SM}} = (249 \pm 59) \times 10^{-11},
\end{equation}
which when compared to eq. (\ref{eq:LW_g-2}) yields the solution $M_{\mu} \simeq 5.2$ TeV. Unfortunately, this is already excluded by the bound in eq. (\ref{eq:mu_bound}). Therefore, LW QED cannot account for the $g-2$ anomaly, at least on its own.

Before we conclude this section, we point out that the results in this section are consistent with those obtained from non-local QED with infinite derivatives \cite{Abu-Ajamieh:2023txh,Abu-Ajamieh:2023syy}. There too, it was found that non-locality leads to charge dequantization and a contribution to the magnetic dipole moment, with similar numerical results. This is hardly surprising, since LW theories can be obtained from the leading-order expansion of non-local theories with infinite derivatives \cite{Biswas:2014yia}
\begin{equation}\label{eq:NL&LW}
    \mathcal{L}_{\text{NL}} = -\frac{1}{2}i\Big[ \overline{\Psi} e^{-\frac{D^{2}}{M^{2}}}(\slashed{D}+m)\Psi + h.c. \Big] \simeq   - i\overline{\Psi}\Big(1 - \frac{D^{2}}{M^{2}} \Big) \Big(\slashed{D}+m\Big)\Psi,
\end{equation}

However, we should notice that although the two theories share many similarities, they are nonetheless fundamentally different. For instance, while LW theories introduce new massive poles, theories with infinite derivatives do not add any new degrees of freedom to the particle spectrum.

\section{Implications of the WGC in LW QED}\label{sec:4}
The WGC \cite{Arkani-Hamed:2006emk} roughly states that "gravity is the weakest force". More technically, it states that for any $U(1)$ gauge group of charge $q$, there must exist a particle of mass $m$ satisfying
\begin{equation}\label{eq:El_WGC}
    m \leq \sqrt{2} q M_{\text{P}},
\end{equation}
where $M_{\text{P}}$ is the reduced Planck scale $= 2.4 \times 10^{18}$ GeV. The main motivation of this conjecture stems from the evaporation of extremal black holes. It can be shown that if a particle satisfying eq. (\ref{eq:El_WGC}) does not exist, then black holes cannot decay, which leads to the issue of remnants \cite{Susskind:1995da}. Eq. (\ref{eq:El_WGC}) is called the electric WGC and it can be used to derive another form called the magnetic WGC when applied to magnetic monopoles.
\begin{equation}\label{eq:Mag_WGC}
    \Lambda  \lesssim q M_{\text{P}},
\end{equation}
where $\Lambda$ is a UV cutoff scale were the Effective Field Theory (EFT) breaks down. The magnetic WGC implies that any gauged $U(1)$ is associated with a natural cutoff scale below the Planck scale where New Physics (NP) comes into play. For example, for the SM QED, the magnetic WGC implies that $\Lambda \lesssim 10^{17}$ GeV, which corresponds to the scale of the heterotic string.

As LW QED leads inevitably to the dequantization of the electric charge, it is worthwhile investigating what implication the WGC has on it. First, let's investigate the implications of the magnetic WGC. It was shown in \cite{Abu-Ajamieh:2024gaw} that the magnetic WGC, combined with the bounds on matter neutrality in eq. (\ref{eq:Nu_Q_limit}), implies that neutrinos are electrically neutral and thus charge is quantized. This was found as follows: LEP searches can be used to set a lower limit on the scale of NP with vector mediators $\Lambda \sim 10$ TeV. This can be translated in to a lower bound on the charge of neutrinos $e \epsilon \gtrsim \Lambda/M_{\text{P}} \approx 1.4 \times 10^{-14}$, which is excluded according to bound in eq. (\ref{eq:Nu_Q_limit}). Another way to look at this is to notice that the bound in eq. (\ref{eq:Nu_Q_limit}) implies that the scale of NP is given by
\begin{equation}\label{eq:NP_scale}
    \Lambda \lesssim \epsilon e M_{\text{P}} = (10^{-21})(0.313)(2.4 \times 10^{18}) \simeq 1 \hspace{1mm} \text{MeV},
\end{equation}
which is clearly excluded! This implies that there is tension between the magnetic WGC and LW QED. However, we should point out that the magnetic WGC was challenged in \cite{Saraswat:2016eaz}, although the author argued that the electric WGC remains valid. In this paper, we choose to be conservative and limit ourselves to the electric WGC, keeping in mind that should the magnetic versions be proven correct, then it would subsequently exclude LW QED and any extensions that lead to charge dequantization. Inserting the bound in eq. (\ref{eq:Nu_Q_limit}) in the electric WGC yields an upper bound on the masses of neutrinos $m_{\nu_{i}} \lesssim 1.08$ MeV. Given the upper bounds on neutrino masses from the PDG \cite{pdg}
\begin{align}
    m_{\nu_{e}} & < 1.1 \hspace{2mm} \text{eV}, \label{eq:pdg1}\\
    m_{\nu_{\mu}} & < 0.19 \hspace{2mm} \text{MeV}, \label{eq:pdg2}\\
    m_{\nu_{\tau}} & < 18.9 \hspace{2mm} \text{MeV}, \label{eq:pdg3}
\end{align}
we can see that while $m_{\nu_{e}}$ and $m_{\nu_{\mu}}$ do indeed satisfy this bound, $m_{\nu_{\tau}}$ might not. This furnishes an experimental test for either LW QED or the WGC: If experiment finds that $m_{\nu_{\tau}} \gtrsim 1.08$ MeV, then this would imply tension between the LW QED and the WGC, which means that either neutrinos are neutral, thereby excluding LW QED, or that the WGC is invalidated if neutrinos are found to have an electric charge.

Eq. (\ref{eq:epsilon}) can be used in the WGC to set bounds on the LW scale of neutrinos. Plugging $\epsilon$ in eq. (\ref{eq:El_WGC}), we find
\begin{equation}
    M_{\nu_{i}} \leq \sqrt{2\sqrt{2}m_{\nu_{i}} e M_{\text{P}}},
\end{equation}
which given the upper bounds on the neutrino masses translates into the following bounds on the neutrinos LW scale
\begin{align}
    M_{\nu_{e}} & \lesssim 47 \hspace{2mm} \text{TeV}, \label{eq:M_nu_e} \\
    M_{\nu_{\mu}} & \lesssim 2.02 \times 10^{4} \hspace{2mm} \text{TeV}, \label{eq:M_nu_mu} \\
    M_{\nu_{\tau}} & \lesssim 2.02 \times 10^{5} \hspace{2mm} \text{TeV} \label{eq:M_nu_tau},
\end{align}
which implies that the neutrinos LW scale could be within the reach of future colliders, such as the proposed 100-TeV collider or muon collider. Comparing these bounds with those on charged lepton in eq. (\ref{eq:e_bound})-(\ref{eq:tau_bound}), we see that in addition to the neutrinos bounds being significantly lower, they constitute \textit{upper} bounds as opposed to the \textit{lower} bounds on the charged leptons. This is quite intriguing, as one would naively expect the LW scales of each lepton-neutrino pair to be related. This seems to suggest that there is tension between the bounds obtained from experiment and those suggested by the WGC, however, nothing conclusive can be drawn, as in principle the LW scales of different particle could indeed be independent of one another.

\section{The Modified Electric Potential and the WGC in LW QED}\label{sec:5}
\begin{figure}[!t]
\centerline{\begin{minipage}{0.75\textwidth}
\centerline{\includegraphics[width=200pt]{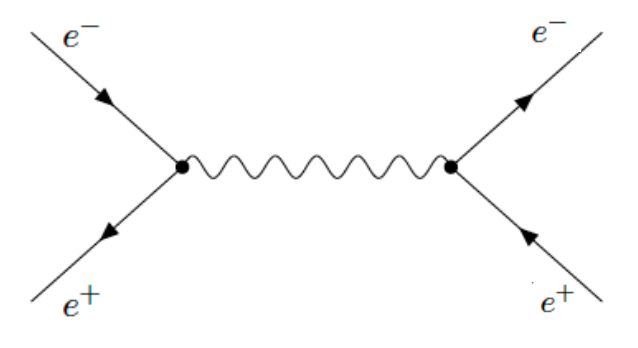}}
\caption{\small S-channel $e^{+}e^{-} \rightarrow e^{+}e^{-}$ scattering.}
\label{fig1}
\end{minipage}}
\end{figure}

Since the photon propagator in LW QED is modified compared to the ordinary case, the electric potential will subsequently be modified. To calculate the LW potential, consider the s-channel scattering $e^{+}e^{-} \rightarrow e^{+}e^{-}$ in LW QED shown in Fig. (\ref{fig1}), the matrix element is expressed as
\begin{equation}\label{eq:ee_scattering}
    i\mathcal{M} = (-ie)^{2} \overline{u}(p_{2})\gamma^{\mu}u(p_{1}) \Bigg[ \frac{-i}{k^{2} - k^{4}/M_{A}^{2}} \Big( \eta_{\mu\nu} -(1-\xi)\frac{k_{\mu}k_{\nu}}{k^{2}} -\xi \frac{k_{\mu}k_{\nu}}{M_{A}^{2}} \Big) \Bigg] \overline{u}(q_{2})\gamma^{\nu}u(q_{1}).
\end{equation}

\begin{figure}[!t]
    \centering
    \subfloat{{\includegraphics[width=210pt]{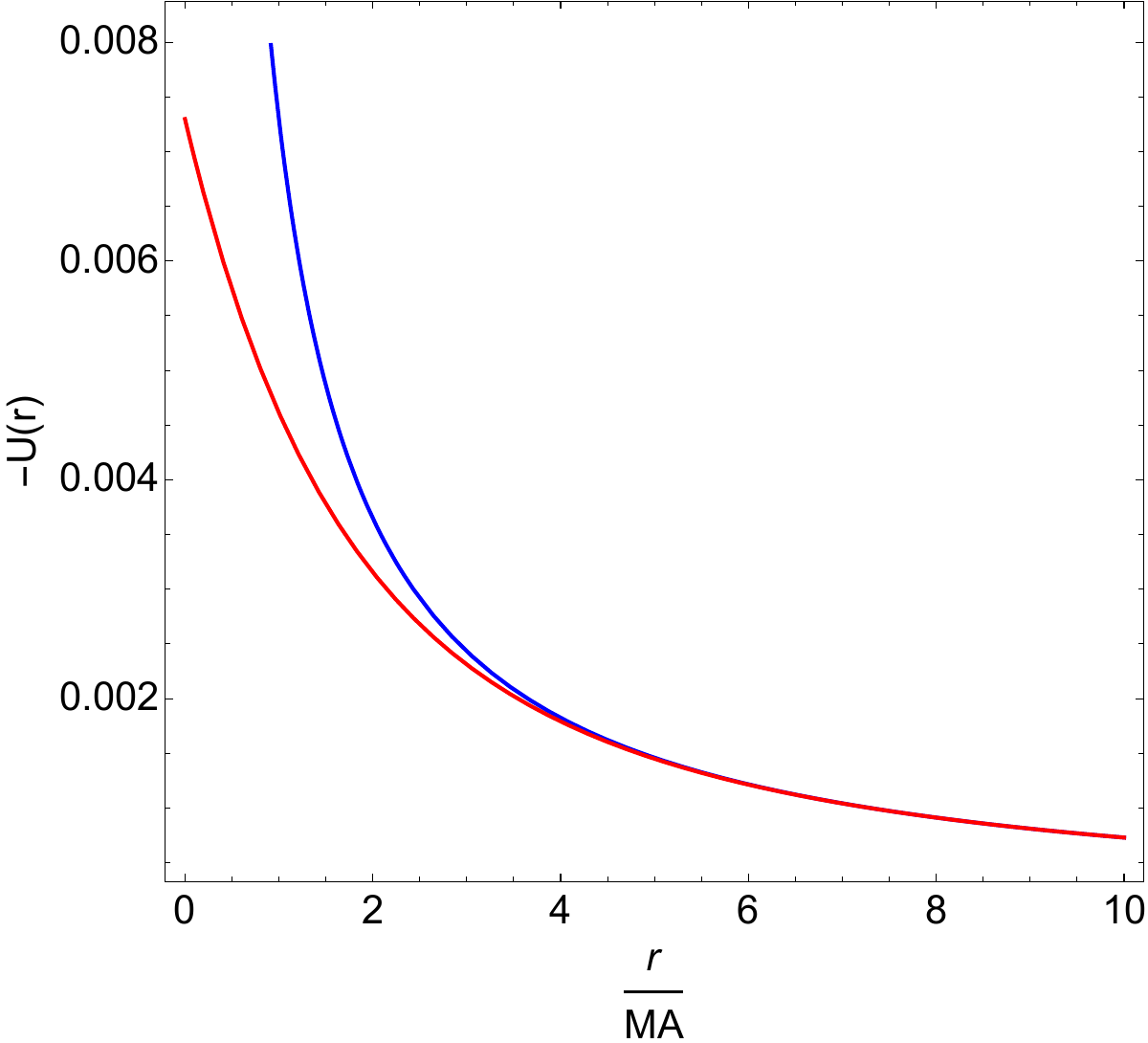} }}%
    \qquad
    \subfloat{{\includegraphics[width=210pt]{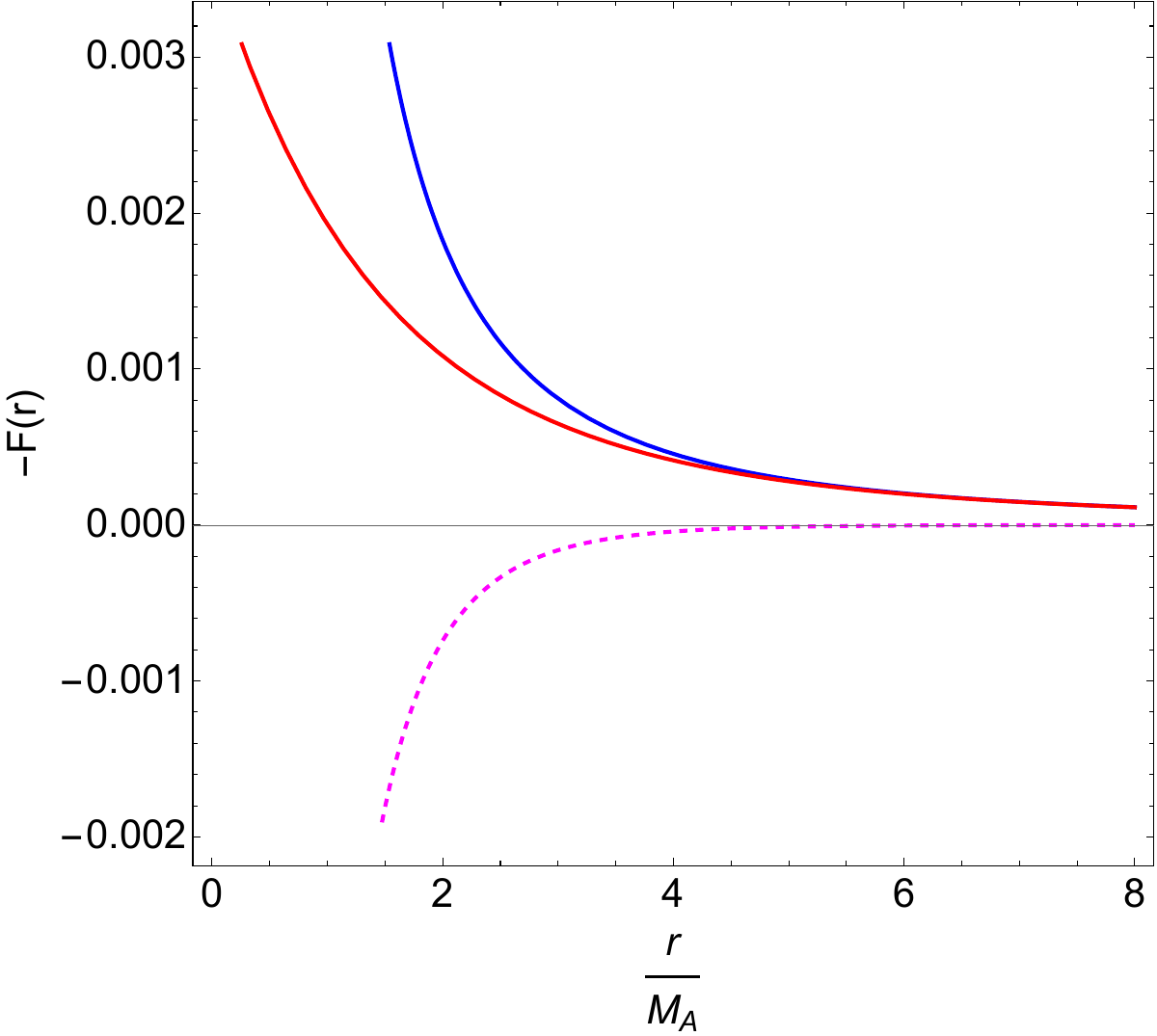} }}%
    \caption{\small (Left): The LW potential (red) vs. the Coulomb potential (blue). (Right): The LW force (red) vs. the Coulomb force (blue). The dashed line is the LW modification term to the Coulomb force. We see that while the LW force/potential approach their Coulomb counterparts in the low energy limit, they nonetheless deviate drastically in the high energy limit. In particular, while the Coulomb force/potential blow up when $r \rightarrow 0$, their LW counterparts remain finite.}
    \label{fig2}
\end{figure}

The electric potential is obtained from the photon propagator at vanishing energy transfer. Notice that in the non-relativistic limit, we have
\begin{align}\label{eq:non_rel_spinors}
    \overline{u}(p_{2})\gamma^{0}u(p_{1}) & \simeq 2m \xi^{\dagger}\xi, \\
    \overline{u}(p_{2})\gamma^{i}u(p_{1}) & \simeq 0,
\end{align}
which implies that only $\mu = \nu = 0$ contributes. Thus, we can find the LW electric potential as follows
\begin{align}\label{eq:LW_potential}
    & U_{\text{LW}}(\bm{k}) = \lim_{k_{0}\rightarrow 0} \Bigg[ \frac{e^{2}}{k_{0}^{2} - |\bm{k}|^{2}- \frac{1}{M_{A}^{2}}(k_{0}^{2} - |\bm{k}|^{2})^{2}} \Big( \eta_{00} -(1-\xi)\frac{k_{0}^{2}}{k^{2}} -\xi \frac{k_{0}^{2}}{M_{A}^{2}} \Big) \Bigg], \nonumber \\   
    \implies & U_{\text{LW}}(r)  =  -\frac{\alpha}{r} \Big(1 - e^{-r M_{A}} \Big)
\end{align}
where in the second step we transformed to position space. This result agrees with the one found in \cite{Turcati:2016aog}. We see that the result is independent of the gauge, and we can also see that when $M_{A} \rightarrow \infty$, we obtain the (ordinary) Coulomb potential. We plot the LW and Coulomb potentials on the left hand side of Fig. (\ref{fig2}). We see from the plot that while the LW and Coulomb potentials are almost identical at low energy, they nonetheless deviate drastically from each other at high energy. In fact, while the Coulomb potential blows up in the limit $r \rightarrow 0$, the LW potential remains finite
\begin{equation}\label{eq:LW_potential_limit}
    \lim_{r \rightarrow 0} U_{\text{LW}}(r) = - \alpha M_{A}.
\end{equation}

This shows how adding the LW high-order operator helps regularize UV divergences. To find the LW electric force, we simply take the derivative of eq. (\ref{eq:LW_potential}) w.r.t. $r$
\begin{equation}\label{eq:LW_force}
    F_{\text{LW}}(r) = -\frac{\alpha}{r^{2}} + \frac{\alpha}{r^{2}}(1+r M_{A}) e^{-r M_{A}},
\end{equation}
where the first term corresponds to the usual Coulomb force, whereas the second is the LW modification. We plot the LW force and the Coulomb force on the right hand side of  Fig. (\ref{fig2}). Similar to the electric potential, the LW force  approaches the Coulomb one in the low energy limit and deviates from it in the high energy limit. In the plot, we explicitly show the LW modification term. This term has the opposite sign of the Coulomb force and is negligible in the low energy regime, but becomes significant in the high energy regime, thereby rendering the LW electric force finite as $r \rightarrow 0$
\begin{equation}\label{eq:LW_Force_lim}
    \lim_{r \rightarrow 0} F_{\text{LW}} = -\frac{1}{2} \alpha M_{A}^{2}.
\end{equation}

In the previous section, we investigated the implications of the WGC on LW QED, however, we should point out that the WGC above was formulated for $U(1)$ gauge theories without higher-order derivatives, whereas LW QED contains higher-order derivatives and thus has a modified force as we saw. Therefore, it is worthwhile asking the question: How is the WGC modified in the LW QED?

To answer this question, we notice that the physical content of the WGC is that the gravitational force is weaker than the force associated with any $U(1)$ gauge theory, i.e.,
\begin{equation}\label{eq:WGC_derivation}
    F_{\text{grav}} = \frac{m^{2}}{8\pi M_{\text{P}}^{2}r^{2}} \leq F_{U(1)} = \frac{q^{2}}{4\pi r^{2}},
\end{equation}
which immediately yields eq. (\ref{eq:El_WGC}). In light of this, we can derive the WGC in LW QED simply by requiring that the gravitational force be weaker than the LW electric force in eq. (\ref{eq:LW_force}). This immediately leads to
\begin{equation}\label{eq:El-LW_WGC}
    m \leq \sqrt{2} q M_{\text{P}} \Big[1-(1+r M_{A}) e^{-r M_{A}} \Big]^{1/2},
\end{equation}
and the magnetic form can be easily read-off
\begin{equation}\label{eq:Mag_LW_WGC}
    \Lambda \lesssim q M_{\text{P}} \Big[1-(1+r M_{A}) e^{-r M_{A}} \Big]^{1/2},
\end{equation}
and it's easy to see that in the limit $M_{A} \rightarrow \infty$, we retrieve the ordinary WGCs in eqs. (\ref{eq:El_WGC}) and (\ref{eq:Mag_WGC}). The ordinary cases are also retrieved in the limit $r \rightarrow \infty$ corresponding to the low-energy limit, whereas in the high energy limit $r \rightarrow 0$, eqs. (\ref{eq:El-LW_WGC}) and (\ref{eq:Mag_LW_WGC}) become\footnote{Notice that more precisely, this expansion is not accurate and only serves as an illustration, as in the region $r \ll M_{A}^{-1}$, LW QED breaks down as an EFT, and the WGC also is modified since the forces are no longer given by the low energy expressions.}
\begin{align}
    m & \leq q r M_{A}M_{\text{P}}, \\
    \Lambda & \lesssim q r M_{A}M_{\text{P}}.
\end{align}

\begin{figure}[!t]
\centerline{\begin{minipage}{0.75\textwidth}
\centerline{\includegraphics[width=250pt]{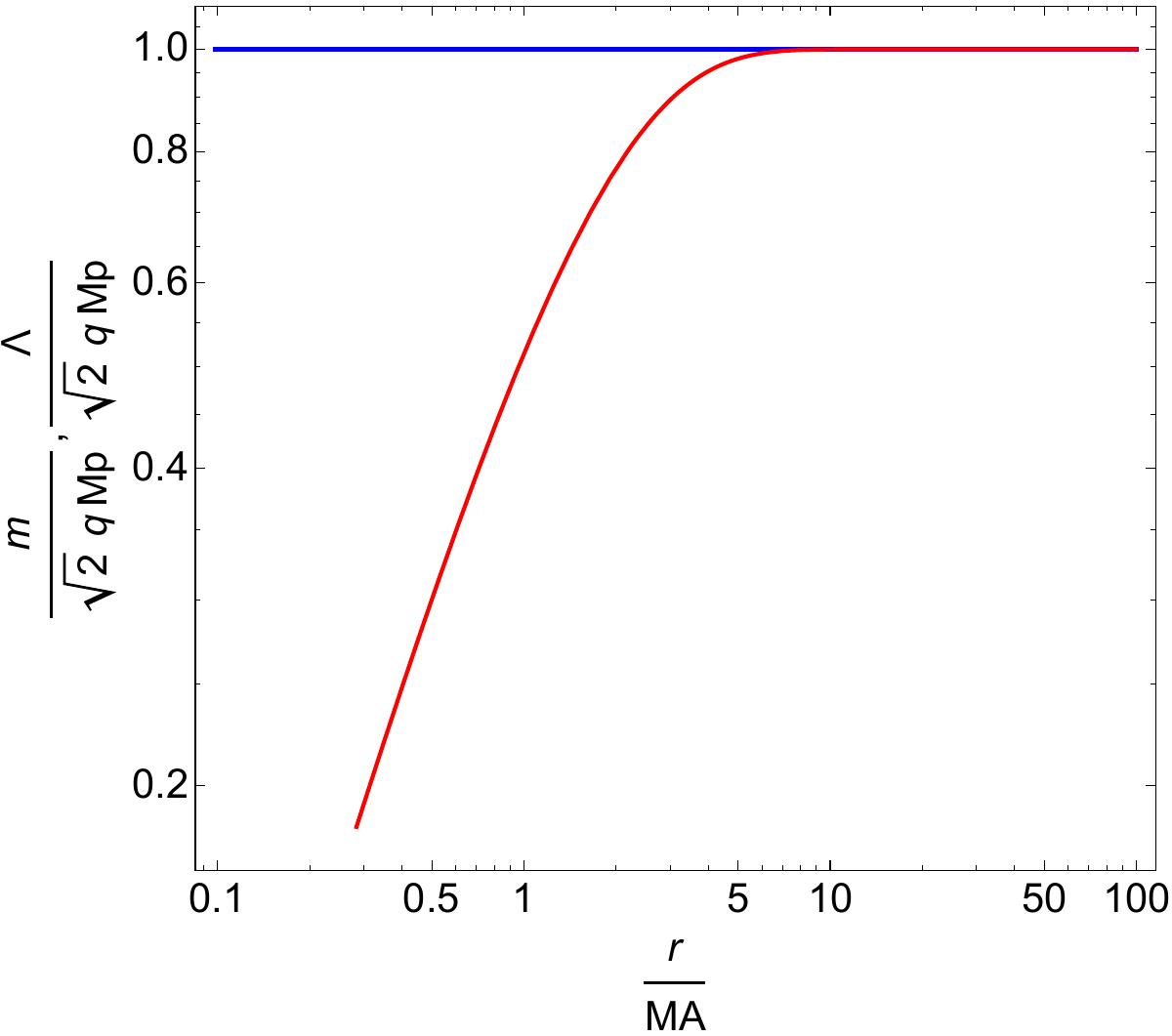}}
\caption{\small The WGC in the ordinary (blue) and LW (red) QED normalized so that the former = 1. The LW WGC yields stronger bounds.}
\label{fig3}
\end{minipage}}
\end{figure}

We plot the LW WGC vs the ordinary one in Fig. (\ref{fig3}). As the plot shows, while both the LW and the ordinary WGCs are essentially identical in the low energy regime, the LW version yields stronger bounds in the high energy regime, i.e., the mass $m$ becomes lighter and the cutoff scale $\Lambda$ becomes lower. Thus, the bounds on NP in the previous section are expected to be even lower. Notice however, that the LW WGC depends on the energy scale itself through $r$, which limits its applicability for practical purposes, as one usually lacks a predefined energy scale to apply the conjecture. Thus to be conservative, we adopt the bounds in the previous section. Nonetheless, eqs. (\ref{eq:El-LW_WGC}) and (\ref{eq:Mag_LW_WGC}) are useful qualitatively, as they inform us that in LW theories, we expect NP at a scale lower than ordinary QED.

\section{Photon Self-energy and The LW Uehling Potential}\label{sec:6}
In this section, we will compute the photon self-energy in LW QED and use the result to calculate the modification to the electric potential (i.e. the Uehling potential) and the Lamb shift. These calculations were done in \cite{Turcati:2016aog} for QED with a higher photon derivative operator, i.e., adding the operator $-\frac{1}{4M_{A}^{2}}F_{\mu\nu} \Box F^{\mu\nu}$, which only includes the higher-order \textit{ordinary} derivative. This is different from our case where we include the higher-order \textit{covariant} derivatives in both the photon and the fermion sectors, which modify the interaction vertex in addition to the propagators. Thus, our results are different from theirs.

\begin{figure}[!t]
\centerline{\begin{minipage}{0.75\textwidth}
\centerline{\includegraphics[width=300pt]{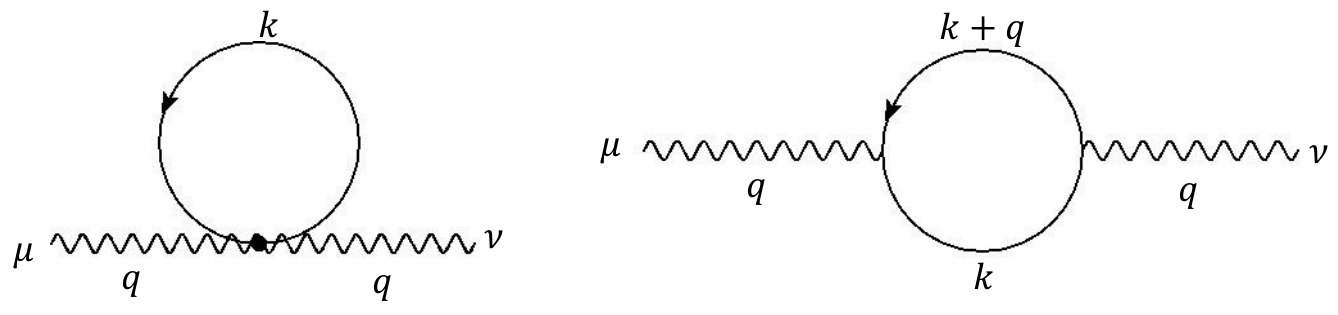}}
\caption{\small Photon self-energy in LW QED. The L.H.S. diagram arises from the expansion of the higher-order covariant derivaives and must be set to vanish.}
\label{fig4}
\end{minipage}}
\end{figure}

To compute the self-energy, we note that the Feynman rules obtained from the higher derivative LW Lagrangian give rise to two diagrams as shown in Fig. (\ref{fig4}). The diagram on the R.H.S. is the same as the photon self-energy we find in ordinary QED, whereas the one on the L.H.S is novel and arises from the  expansion of the covariant derivatives in the LW term. A quick calculation of this diagram in the higher-derivative formalism shows that the amplitude $\sim M_Q^2$, and since we need to restore the SM results in the limit $M_Q \rightarrow \infty$, we need to set the renormalization condition such that this diagram vanishes. Thus, only the second diagram contributes to the self-energy.

As shown in Section \ref{sec:2}, it is more convenient to use the auxiliary field formalism to calculate loops. The auxiliary field formalism introduces two LW Dirac fermions that we label as $Q$. Thus, the second diagram actually represents 3 contributions: One with the ordinary fermion running in the loop, and the other 2 with the LW fermions running in the loop. The contribution of each diagram is given by

\begin{equation}\label{eq:SE_AF_1}
i\Pi_i^{\mu\nu} = - e^{2}\int \frac{d^{4}k}{(2\pi)^{4}}\text{Tr}\Bigg[ \frac{\gamma^{\mu}(\slashed{k}+\slashed{q}+M_i)\gamma^{\nu}(\slashed{k}+M_i)}{((k+q)^{2}-M_i^{2})(k^{2}-M_i^{2})} \Bigg],
\end{equation}
where $M_1 = m$ is the mass of the ordinary fermion, $M_2 =M_Q$ and $M_3 =-M_Q$ are the masses of the LW fermions, and it's easy to see that gauge-invariance is preserved. All 3 contributions modify the photon self-energy, which can be expressed as
\begin{equation}
   i\Pi^{\mu\nu}(p) = i\Pi_a^{\mu\nu}(p) + i\Pi_b^{\mu\nu}(p) + i\Pi_c^{\mu\nu}(p)  =(p^2g^{\mu\nu}-p^\mu p^\nu)i\Pi_2(p),
\end{equation}
where the renormalized 2-point function is given by
\begin{align}
\label{eq:loglog}
    {\widehat \Pi}_2(p) & =\Pi_2(p)-\Pi_2(0), \nonumber\\
    & =-\frac{2\alpha}{\pi}
    \int_0^1dxx(1-x)\left[\log\left(\frac{m^2}{m^2-x(1-x)p^2}\right)+2\log\left(\frac{M_Q^{2}}{M_Q^{2}-x(1-x)p^2}\right)\right].
\end{align}

Notice that the 2 LW fermions yield identical contributions, and that in the limit $M_Q\rightarrow \infty$, the second term vanishes and the ordinary QED result is restored. As in the ordinary QED case, ${\widehat \Pi}_2(p)$ modifies the electric potential in the non-relativistic limit yielding the Uehling potential. However, here we should pay attention to the fact that we need to use the LW electric potential in eq. (\ref{eq:LW_potential}) to calculate the LW Uehling potential, as opposed to using the Coulomb potential when calculating the SM Uehling potential. Thus, the LW Uehling potential is given by
\begin{equation}\label{eq:Uheling1}
    U({\bm r})=\int\frac{d^3k}{(2\pi)^3}e^{i{\bm k}\cdot{\bm r}}\frac{-e^2}{|{\bm k}|^2\left(1+|{\bm k}|^2/M_A^2\right)[1-{\widehat \Pi}_2(-|{\bm k}|^2)]}.
\end{equation}
As a first approximation, let's evaluate the integral in the limit of small momentum $|p|^2\ll m^2$. In this limit, eq. (\ref{eq:Uheling1}) becomes
\begin{eqnarray}\label{eq:Uehling2}
    U({\bm r}) & = &\int\frac{d^3k}{(2\pi)^3}e^{i{\bm k}\cdot{\bm r}}\frac{-e^2}{|{\bm k}|^2\left(1+|{\bm k}|^2/M_A^2\right) \Big[1+\frac{\alpha}{15\pi}|{\bm k}|^2\left(\frac{1}{m^2}+\frac{2}{M_Q^2}\right) \Big]},
    \nonumber \\
    &\approx&-\frac{\alpha}{r}-\frac{\alpha^2}{15\pi}M_A^2\left(\frac{1}{m^2}+\frac{2}{M_Q^2}-\frac{15\pi}{M_A^2\alpha}\right)\frac{e^{-M_Ar}}{r}, \nonumber \\
    &\approx & U_{\text{LW}}(r)-\frac{\alpha^2}{15\pi}M_A^2\left(\frac{1}{m^2}+\frac{2}{M_Q^2}\right)\frac{e^{-M_Ar}}{r},
\end{eqnarray}
where $U_{\text{LW}}(r)$ is the LW QED potential given in  eq.(\ref{eq:LW_potential}). Now that we have the Uehling potential in LW QED, it is possible to calculate the Lamb shift, which is simply the shift in the energy level of the 2S state of the hydrogen atom due to the modification in the Coulomb potential. In LW QED the Lamb shift is given by
\begin{align}\label{eq:LS}
   \Delta E_{\text{LW}} &= -\frac{\alpha^2}{15\pi}M_A^2\left(\frac{1}{m^2}+\frac{2}{M_Q^2}-\frac{15\pi}{M_A^2\alpha}\right)\frac{m\alpha}{4}\frac{1+\frac{2M_A^2}{\alpha^2m^2}}{\left(1+\frac{M_A}{\alpha m}\right)^4}, \nonumber\\
    & =\frac{1}{2\alpha^2}
   \left(\frac{M_A^2}{m^2}+2\frac{M_A^2}{M_Q^2}-\alpha\right)\frac{1+\frac{2M_A^2}{\alpha^2m^2}}{\left(1+\frac{M_A}{\alpha m}\right)^4} \times  \Delta E_{\text{LS}},
\end{align}
where $\Delta E_{LS}=-m\alpha^5/30\pi$ is the Lamb shift in ordinary QED. Assuming $M_A\approx M_Q$, we can use eq. (\ref{eq:LS}) to set bounds on the LW scale $M_{A}$ from the bounds on the Lamb shift. For the transition $2S_{1/2}-2P_{1/2}$, the difference between the theoretical and experimental measurement of the Lamb shift is given by \cite{Eides:2000xc}
\begin{equation}
    |\Delta E^{\text{Th}} -\Delta E^{\text{Exp}}| = |1057.833 - 1057.845| = 0.012 \hspace{2mm} \text{MHz},
\end{equation}
which translates to the lower bound
\begin{equation}
    M_A \gtrsim 1 \hspace{2mm} \text{GeV}.
\end{equation}

We see that the bound obtained from the Lamb shift is quite low, which is consistent with what was found in the case of non-local QED \cite{Abu-Ajamieh:2023txh}. Notice that the results in eq. (\ref{eq:Uehling2}) is only an approximation. The exact expression for the Uehling potential is given by
\begin{equation}
    U({\bm r})=\frac{ie^2}{(2\pi)^2r}\int_{-\infty}^\infty dQ\frac{Qe^{iQr}}{(Q^2+\mu^2)\left(1+\frac{Q^2}{M_A^2}\right)}[1+{\widehat \Pi}_2(-Q^2)],
\end{equation}
where have defined $Q\equiv|{\bm p}|$ and we've added a regulator $\mu$ to the Coulomb potential. Notice that to find the exact potential, we need to calculate the contribution of the branch cut, which requires the evaluation of the imaginary part of ${\widehat \Pi}_2(-Q^2)$
\begin{align}
    \text{Im}{\widehat \Pi}_2(q^2\pm i\epsilon) & =   -\frac{2\alpha}{\pi}(\pm\pi)\int_{\frac{1}{2}-\frac{\beta}{2}}^{\frac{1}{2}+\frac{\beta}{2}}dx x(1-x)
    -\frac{4\alpha}{\pi}(\pm\pi)\int_{-\infty}^{\frac{1}{2}-\frac{\beta_1}{2}}dx x (1-x)\nonumber\\
    & -\frac{4\alpha}{\pi}(\pm\pi)\int_{\frac{1}{2}+\frac{\beta_1}{2}}^{\infty}dx x (1-x),
\end{align}
where $\beta=\sqrt{1-4m^2/p^2}$ and $\beta_1=\sqrt{1+4M_Q^2/p^2}$. The last two integrals are out of the integration range since $0\le x\le 1$, and thus do not contribute. Therefore, the integral yields
\begin{equation}
    \text{Im}{\widehat \Pi}_2(p^2+i\epsilon)=\mp\frac{\alpha}{3}\sqrt{1-\frac{4m^2}{p^2}}\left(1+\frac{2m^2}{p^2}\right),
\end{equation}
which is identical to the SM case, and we can see the branch cut at $p=2im$. Now we can use this result to find the modification to the electric potential due to the self-energy correction
\begin{equation}
    \delta U(r)=-\frac{2\alpha}{\pi r}\int_{2m}^\infty dp\frac{e^{ipr}}{p\left(1+\frac{p^2}{M_A^2}\right)}\frac{\alpha}{3}\sqrt{1-\frac{4m^2}{p^2}}\left(1+\frac{2m^2}{p^2}\right).
\end{equation}

This integral is dominated near $p\approx 2m$. Therefore, after evaluating the the integral in this region and adding the result to the results in eq. (\ref{eq:LW_potential}), we finally arrive at the final Uehling potential in LW QED
\begin{equation}
\label{eq:FP}
    U(r)=-\frac{\alpha}{r} \Big(1 - e^{-r M_{A}} \Big) -\frac{\alpha^2}{4\sqrt{\pi}r}\frac{1}{(1+\frac{4m^2}{M_A^2})}\frac{e^{-2mr}}{(mr)^{\frac{3}{2}}},
\end{equation}
and we see that in the limit $M_{A} \rightarrow \infty$, the SM Uehling potential is restored. Notice that apart from the tree-level modification, the LW correction to the Uehling potential is very small $\sim m^2/M_A^2$ given the bounds on $M_A$. 

\begin{figure}[!t]
\centerline{\begin{minipage}{0.75\textwidth}
\centerline{\includegraphics[width=300pt]{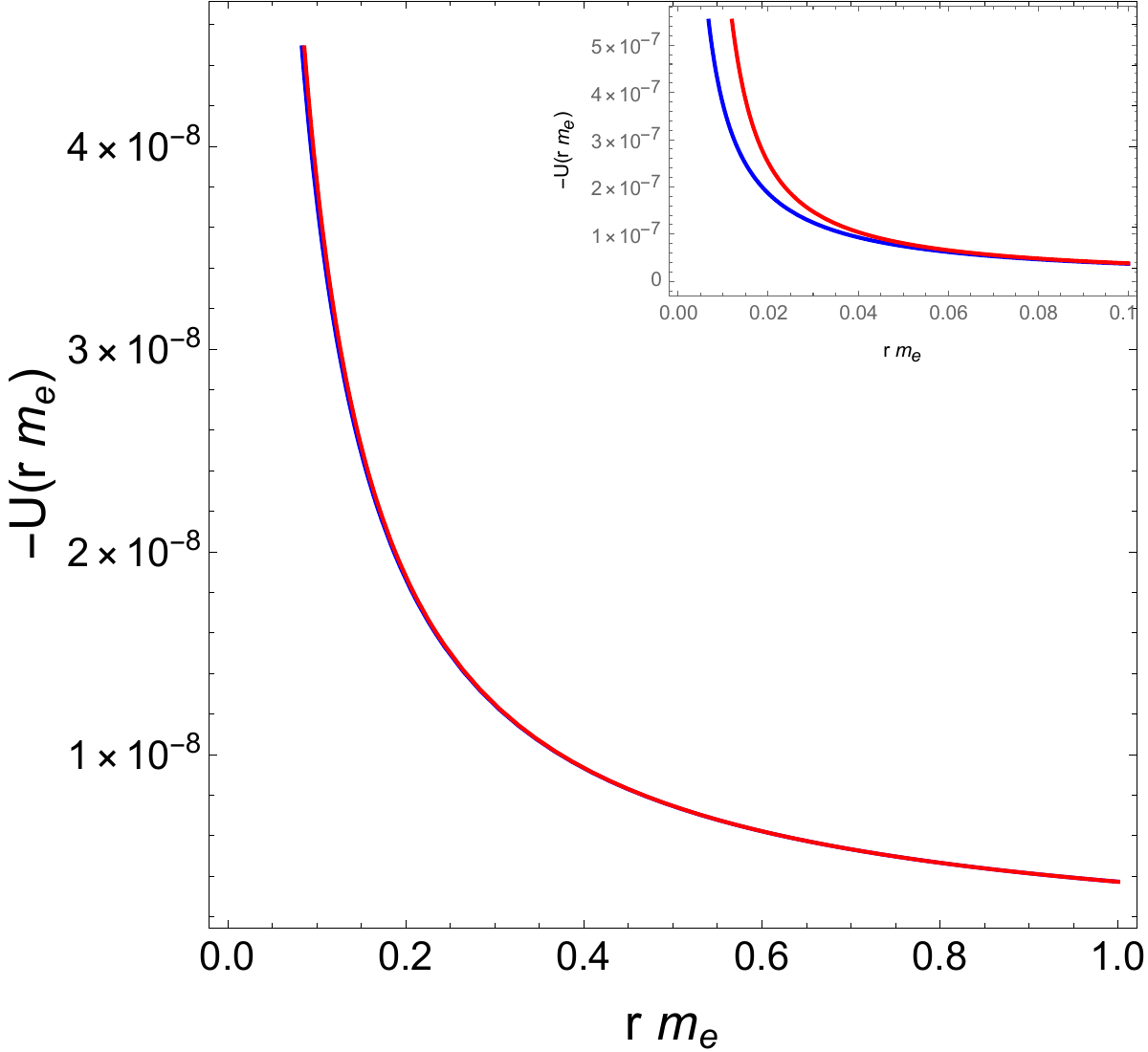}}
\caption{\small The LW Uehling potential (red) given in eq. (\ref{eq:FP}) vs. the Coulomb potential (blue). The two potentials are very close to one another, but deviate at high energy.}
\label{fig5}
\end{minipage}}
\end{figure}

We plot the LW Uehling potential in Figure (\ref{fig5}), together with the Coulomb potential, where we see the usual behavior: The two potentials agree in the low energy limit and deviate from one another in the high energy limit. This is due to the usual screening effect that becomes more prominent at high energy.

Before we conclude this section, we should point out that including the LW corrections (such as EW corrections) in the above calculation is probably significant, however, this is beyond the scope of this paper.

\section{Conclusions}\label{sec:7}
In this paper, we studied some phenomenological aspects of LW QED. In particular, we showed that in LW QED the electric charge becomes dequantized and showed that the only way to accommodate this charge dequantization is through a flavor-dependent LW scale. We used the bounds on the neutrino charge to set lower bounds on these LW scales for the SM charged fermions, which were found to range between $\sim O(10^{4})$ TeV for the electron to $\sim O(10^{10})$ TeV for the top quark. We also showed that LW QED will yield an additional contribution to the magnetic dipole moment, however, this contribution cannot alone explain the muon $g-2$ anomaly.

We studied the implications of the WGC on LW QED and found that there is tension between the magnetic WGC and LW QED, as the magnetic WGC implies charge quantization. On the other hand, ignoring the magnetic WGC and confining ourselves to the electric one, we put upper limits on the LW scale for neutrinos and found that it could be as low as $\sim 47$ TeV, which could be within the reach of future colliders.

We calculated the electric force and potential in LW QED and found that while they agree with their Coulomb counterpart at low energy, they nonetheless deviate drastically from them and remain finite at high energy, which explains how UV divergences are regularized in LW QED. We also used the modified electric force to reformulate the WGC in LW QED and found that compared to the ordinary WGC, it yields stronger bounds.

Finally, we evaluated the photon's self-energy and used the result to calculate the Lamb shift and the Uehling potential in LW QED. We found that the Lamb shift leads to a weak limit on the LW scale $M_A$, and found that the modification of Uehling potential due to the LW masses $M_A$ and $M_Q$ is small. We point that precision QED measurements can be used to probe LW QED and could potentially exclude it.

\section*{Acknowledgments}
The work of FA is supported by the C.V. Raman fellowship from CHEP at IISc. 
\appendix

\section{The Auxiliary Field Formalism}
In this paper, we have used the auxiliary field formalism developed in \cite{Grinstein:2007mp} for loop calculation. In this formalism, the Lee Wick Lagrangian for fermions can be expressed as 
\begin{align}
   \mathcal{L}_{AF}={\bar\Psi}(i{\slashed D-m})\Psi
    +\bar{\tilde{\Psi}}(i\slashed {D}-M_Q)\tilde{\Psi}
    +\bar{\hat{\Psi}} (i\slashed{D}+M_Q)\hat{\Psi},
\end{align} 
where $\tilde{\Psi}$ and $\hat{\Psi}$ are the auxiliary fermionic fields. The Feynman rules for the above Lagrangian are given by
\begin{align} 
\langle \bar{\Psi}\Psi\rangle &= \frac{i(\slashed{p}+m)}{p^2-m^2},\\
\langle \bar{\tilde{\Psi}}\tilde{\Psi}\rangle & = \frac{i(\slashed{p}+M_Q)}{(p^2-M_Q)^2},\\
\langle \bar{\hat{\Psi}}\hat{\Psi}\rangle & =\frac{i(\slashed{p}-M_Q)}{(p^2-M_Q)^2},\\
\langle \bar{\Psi}\Psi A^{\mu}\rangle & = ie\gamma^{\mu},
\end{align}
\begin{align} 
\langle \bar{\tilde{\Psi}}\tilde{\Psi} A^{\mu}\rangle & =-ie\gamma^{\mu},\\
\langle \bar{\hat{\Psi}}\hat{\Psi} A^{\mu}\rangle & = ie\gamma^{\mu}.
\end{align}


\begin{thebibliography}{10}


\bibitem{Lee:1969fy}
T.~D.~Lee and G.~C.~Wick,
``Negative Metric and the Unitarity of the S Matrix,''
Nucl. Phys. B \textbf{9}, 209-243 (1969)

\bibitem{Lee:1970iw}
T.~D.~Lee and G.~C.~Wick,
``Finite Theory of Quantum Electrodynamics,''
Phys. Rev. D \textbf{2}, 1033-1048 (1970)

\bibitem{Coleman}
S. Coleman, in Erice 1969: Ettore Majorana School on Subnuclear Phenomena edited by A. Zicchici (Academic Press, New York, 1970), p. 282.

\bibitem{Cutkosky:1969fq}
R.~E.~Cutkosky, P.~V.~Landshoff, D.~I.~Olive and J.~C.~Polkinghorne,
``A non-analytic S matrix,''
Nucl. Phys. B \textbf{12}, 281-300 (1969)

\bibitem{Grinstein:2007mp}
B.~Grinstein, D.~O'Connell and M.~B.~Wise,
``The Lee-Wick standard model,''
Phys. Rev. D \textbf{77}, 025012 (2008)
\arXivold{0704.1845}{hep-ph}.

\bibitem{Frasca:2022duz}
M.~Frasca, A.~Ghoshal and A.~S.~Koshelev,
``Non-perturbative Lee-Wick gauge theory: Towards Confinement \& RGE with strong couplings,''
Class. Quant. Grav. \textbf{41}, no.1, 015014 (2024)
\arXivold{2202.09578}{hep-ph}.

\bibitem{Modesto:2016ofr}
L.~Modesto,
``Super-renormalizable or finite Lee\textendash{}Wick quantum gravity,''
Nucl. Phys. B \textbf{909} (2016), 584-606
\arXivold{1602.02421}{hep-th}.

\bibitem{Okun:1983vw}
L.~B.~Okun, M.~B.~Voloshin and V.~I.~Zakharov,
``ELECTRICAL NEUTRALITY OF ATOMS AND GRAND UNIFICATION MODELS,''
Phys. Lett. B \textbf{138}, 115-120 (1984)

\bibitem{Holdom:1985ag}
B.~Holdom,
``Two U(1)'s and Epsilon Charge Shifts,''
Phys. Lett. B \textbf{166}, 196-198 (1986)

\bibitem{Foot:1990uf}
R.~Foot, G.~C.~Joshi, H.~Lew and R.~R.~Volkas,
``Charge quantization in the standard model and some of its extensions,''
Mod. Phys. Lett. A \textbf{5}, 2721-2732 (1990)

\bibitem{Foot:1992ui}
R.~Foot, H.~Lew and R.~R.~Volkas,
``Electric charge quantization,''
J. Phys. G \textbf{19}, 361-372 (1993)
[erratum: J. Phys. G \textbf{19}, 1067 (1993)]
\arXivold{hep-ph/9209259}{hep-ph}.

\bibitem{Capolupo:2022awe}
A.~Capolupo, G.~Lambiase and A.~Quaranta,
``Muon g-2 anomaly and non-locality,''
Phys. Lett. B \textbf{829}, 137128 (2022)
\arXivold{2206.06037}{hep-ph}.

\bibitem{Abu-Ajamieh:2023roj}
F.~Abu-Ajamieh, P.~Chattopadhyay, A.~Ghoshal and N.~Okada,
``Anomalies in string-inspired nonlocal extensions of QED,''
Phys. Rev. D \textbf{109}, no.7, 076013 (2024)
\arXivold{2307.01589}{hep-th}.

\bibitem{Chattopadhyay:2023nbj}
P.~Chattopadhyay and F.~Nortier,
``Ghost-free Electroweak Symmetry Breaking with Weakly Nonlocal Interactions,''
Acta Phys. Polon. B \textbf{55} (2024) no.8, 8-A2
\arXivold{2311.08311}{hep-ph}.

\bibitem{Turcati:2014eaa}
R.~Turcati and M.~J.~Neves,
``Probing features of the Lee-Wick quantum electrodynamics,''
Adv. High Energy Phys. \textbf{2014}, 153953 (2014)
\arXivold{1403.3132}{hep-th}.

\bibitem{Turcati:2016aog}
R.~Turcati and M.~J.~Neves,
``Complex-mass shell renormalization of the higher-derivative electrodynamics,''
Eur. Phys. J. C \textbf{76}, no.8, 456 (2016)
\arXiv{1601.07218}{hep-th}.

\bibitem{Babu:1989tq}
K.~S.~Babu and R.~N.~Mohapatra,
``Is There a Connection Between Quantization of Electric Charge and a ana Neutrino?,''
Phys. Rev. Lett. \textbf{63}, 938 (1989)

\bibitem{Babu:1989ex}
K.~S.~Babu and R.~N.~Mohapatra,
``Quantization of Electric Charge From Anomaly Constraints and a ana Neutrino,''
Phys. Rev. D \textbf{41}, 271 (1990)

\bibitem{Giunti:2014ixa}
C.~Giunti and A.~Studenikin,
``Neutrino electromagnetic interactions: a window to new physics,''
Rev. Mod. Phys. \textbf{87}, 531 (2015)
\arXivold{1403.6344}{hep-ph}.

\bibitem{Raffelt:1999gv}
G.~G.~Raffelt,
``Limits on neutrino electromagnetic properties: An update,''
Phys. Rept. \textbf{320}, 319-327 (1999)

\bibitem{Aoyama:2020ynm}
T.~Aoyama, N.~Asmussen, M.~Benayoun, J.~Bijnens, T.~Blum, M.~Bruno, I.~Caprini, C.~M.~Carloni Calame, M.~C\`e and G.~Colangelo, \textit{et al.}
``The anomalous magnetic moment of the muon in the Standard Model,''
Phys. Rept. \textbf{887}, 1-166 (2020)
\arXivold{2006.04822}{hep-ph}.

\bibitem{Muong-2:2006rrc}
G.~W.~Bennett \textit{et al.} [Muon g-2],
``Final Report of the Muon E821 Anomalous Magnetic Moment Measurement at BNL,''
Phys. Rev. D \textbf{73}, 072003 (2006)
\arXivold{hep-ex/0602035}{hep-ex}.

\bibitem{Muong-2:2021ojo}
B.~Abi \textit{et al.} [Muon g-2],
``Measurement of the Positive Muon Anomalous Magnetic Moment to 0.46 ppm,''
Phys. Rev. Lett. \textbf{126}, no.14, 141801 (2021)
\arXivold{2104.03281}{hep-ex}.

\bibitem{Muong-2:2021ovs}
T.~Albahri \textit{et al.} [Muon g-2],
``Magnetic-field measurement and analysis for the Muon $g-2$ Experiment at Fermilab,''
Phys. Rev. A \textbf{103}, no.4, 042208 (2021)
\arXivold{2104.03201}{hep-ex}.

\bibitem{Muong-2:2021vma}
T.~Albahri \textit{et al.} [Muon g-2],
``Measurement of the anomalous precession frequency of the muon in the Fermilab Muon $g-2$ Experiment,''
Phys. Rev. D \textbf{103}, no.7, 072002 (2021)
\arXivold{2104.03247}{hep-ex}.

\bibitem{Muong-2:2023cdq}
D.~P.~Aguillard \textit{et al.} [Muon g-2],
``Measurement of the Positive Muon Anomalous Magnetic Moment to 0.20 ppm,''
\arXivold{2308.06230}{hep-ex}.

\bibitem{Borsanyi:2020mff}
S.~Borsanyi, Z.~Fodor, J.~N.~Guenther, C.~Hoelbling, S.~D.~Katz, L.~Lellouch, T.~Lippert, K.~Miura, L.~Parato and K.~K.~Szabo, \textit{et al.}
``Leading hadronic contribution to the muon magnetic moment from lattice QCD,''
Nature \textbf{593}, no.7857, 51-55 (2021)
\arXivold{2002.12347}{hep-lat}.

\bibitem{Ce:2022kxy}
M.~C\`e, A.~G\'erardin, G.~von Hippel, R.~J.~Hudspith, S.~Kuberski, H.~B.~Meyer, K.~Miura, D.~Mohler, K.~Ottnad and P.~Srijit, \textit{et al.}
``Window observable for the hadronic vacuum polarization contribution to the muon g-2 from lattice QCD,''
Phys. Rev. D \textbf{106}, no.11, 114502 (2022)
\arXivold{2206.06582}{hep-lat}.

\bibitem{ExtendedTwistedMass:2022jpw}
C.~Alexandrou \textit{et al.} [Extended Twisted Mass],
``Lattice calculation of the short and intermediate time-distance hadronic vacuum polarization contributions to the muon magnetic moment using twisted-mass fermions,''
Phys. Rev. D \textbf{107}, no.7, 074506 (2023)
\arXivold{2206.15084}{hep-lat}.

\bibitem{Abu-Ajamieh:2022nmt}
F.~Abu-Ajamieh and S.~K.~Vempati,
``Can the Higgs Still Account for the g-2 Anomaly?,''
\arXivold{2209.10898}{hep-ph}.

\bibitem{Abu-Ajamieh:2023qvh}
F.~Abu-Ajamieh, M.~Frasca and S.~K.~Vempati,
``Flavor Violating Di- Higgs Coupling,''
\arXivold{2305.17362}{hep-ph}.

\bibitem{Abu-Ajamieh:2023txh}
F.~Abu-Ajamieh, N.~Okada and S.~K.~Vempati,
``Corrected calculation for the non-local solution to the $g -2$ anomaly and novel results in non-local QED,''
JHEP \textbf{01}, 015 (2024)
\arXivold{2309.08417}{hep-ph}.

\bibitem{Abu-Ajamieh:2023syy}
F.~Abu-Ajamieh and S.~K.~Vempati,
``A Proposed Renormalization Scheme for Non-local QFTs and Application to the Hierarchy Problem,''
\arXivold{2304.07965}{hep-th}.

\bibitem{Biswas:2014yia}
T.~Biswas and N.~Okada,
``Towards LHC physics with nonlocal Standard Model,''
Nucl. Phys. B \textbf{898}, 113-131 (2015)
\arXivold{1407.3331}{hep-ph}.

\bibitem{Arkani-Hamed:2006emk}
N.~Arkani-Hamed, L.~Motl, A.~Nicolis and C.~Vafa,
``The String landscape, black holes and gravity as the weakest force,''
JHEP \textbf{06}, 060 (2007)
\arXivold{hep-th/0601001}{hep-th}.


\bibitem{Susskind:1995da}
L.~Susskind,
``Trouble for remnants,''
\arXivold{hep-th/9501106}{hep-th}.

\bibitem{Abu-Ajamieh:2024gaw}
F.~Abu-Ajamieh, N.~Okada and S.~K.~Vempati,
``Implications of the Weak Gravity Conjecture on Charge, Kinetic Mixing, the Photon Mass, and More,''
\arXivold{2401.10792}{hep-ph}.


\bibitem{Saraswat:2016eaz}
P.~Saraswat,
``Weak gravity conjecture and effective field theory,''
Phys. Rev. D \textbf{95}, no.2, 025013 (2017)
\arXivold{1608.06951}{hep-th}.

\bibitem{pdg}
P.A. Zyla et al. (Particle Data Group), Prog. Theor. Exp. Phys. 2020, 083C01 (2020)

\bibitem{Williams:1971ms}
E.~R.~Williams, J.~E.~Faller and H.~A.~Hill,
``New experimental test of Coulomb's law: A Laboratory upper limit on the photon rest mass,''
Phys. Rev. Lett. \textbf{26}, 721-724 (1971)

\bibitem{Eides:2000xc}
M.~I.~Eides, H.~Grotch and V.~A.~Shelyuto,
``Theory of light hydrogen - like atoms,''
Phys. Rept. \textbf{342}, 63-261 (2001)
\arXivold{hep-ph/0002158}{hep-ph}.

\end{thebibliography}
\end{document}